\documentclass[aps,prb,twocolumn,showpacs,amsmath]{revtex4}
\usepackage{graphicx}
\usepackage{bm}

\begin{document}
\title{Effect of nearby Pearl vortices upon the $I_c$ vs $B$ characteristics of planar Josephson junctions in thin and narrow superconducting strips}

\author{John R.\ Clem}
\affiliation{%
   Ames Laboratory and Department of Physics and Astronomy, \\
   Iowa State University, Ames, Iowa, 50011--3160}

\date{\today}

\begin{abstract} 
In this paper I show how to calculate the effect of a nearby Pearl vortex or antivortex upon the critical current $I_c(B)$ when a perpendicular magnetic induction $B$ is applied to a planar Josephson junction in a long, thin  superconducting strip of width $W$ much less than the Pearl length $\Lambda = 2\lambda^2/d$, where $\lambda$ is the London penetration depth and $d$ is the thickness ($d < \lambda$).   The theoretical results provide a qualitative explanation of unusual features recently observed experimentally by Golod {\it et al.}\cite{Golod10} in a device with a similar geometry.
\end{abstract}


\maketitle

\section{\label{intro}Introduction}

Golod et al.\cite{Golod10} recently reported the use of a planar Nb-CuNi-Nb Josephson junction of length $W =$ 3.8 $\mu$m to detect the presence of a nearby  Abrikosov vortex.  A hole of diameter $\sim$30 nm was fabricated in the Nb film at a distance 0.29 $\mu$m from the center of the junction.  The hole could be used to trap a vortex, which carries magnetic flux $\phi_0 = h/2e$ in the same direction as a positive applied magnetic induction B, or an antivortex, which carries $\phi_0$ in the opposite direction.  Without a vortex or antivortex in the hole, the Josephson critical current $I_c(B)$ vs $B$ exhibited a central maximum with secondary peaks roughly resembling the familiar Fraunhofer single-slit diffraction pattern.  However, when an antivortex was trapped in the hole, (i) the central maximum was replaced by a minimum, (ii) the $I_c(B)$ pattern was shifted by $\Delta \Phi \approx \phi_0/2$, (iii) an approximate doubling of the periodicity appeared on one side of the pattern, leading to a clear left-right asymmetry, and (iv) when a vortex was trapped in the hole, the $I_c(B)$ pattern was the mirror image of that for an antivortex.

\begin{figure}
\includegraphics[width=7cm]{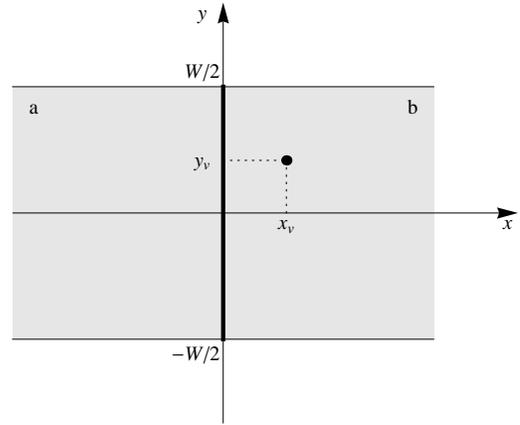}
\caption{Considered here is a long, thin superconducting strip of width $W$ with a planar Josephson junction of width $d_i$  (bold line) at $x = 0$ separating the two halves a and b.  Current leads (not shown) symmetrically feed current $I$ to the sample along the $x$ direction. A magnetic induction $B$ can be applied in the $z$ direction.  A possible vortex position on side b is shown by the black point with coordinates $(x_v,y_v)$. }
\label{sample}
\end{figure} 

To calculate $I_c(B)$ vs $B$ using the exact geometry and material properties used in Ref.\ \onlinecite{Golod10} would be a very difficult numerical problem.  Instead, in this paper I consider a simpler geometry (see Fig.\ \ref{sample}) and solve for $I_c(B)$ vs $B$ in the presence of a nearby vortex or antivortex in the limit that the strip width  $W$  is much less than the Pearl length,\cite{Pearl64} $\Lambda = 2 \lambda^2/d,$  where $\lambda$ is the London penetration depth and $d$ is the strip thickness ($d < \lambda$).  This assumption affords two important simplifications.  An applied magnetic induction $\bm B = \hat z B$  induces screening currents in the film, but when $W \ll \Lambda$, the self-field generated by the screening currents can be neglected.\cite{Moshe08,Clem10}  Moreover, in this limit a vortex in the strip is best described as a Pearl vortex,\cite{Pearl64} whose properties are totally dominated by the $1/r$ sheet-current density circulating around the vortex core generated by the gradient of the order parameter's phase; within a distance $r \ll \Lambda$ from the vortex core the vortex's self-field can be neglected.

Various studies have shown that there is a nonlocal relationship between the Josephson-current distribution in the vicinity of a Josephson vortex core and the magnetic field these currents generate,\cite{Ivanchenko90,Gurevich92,Mints94,Ivanchenko95,Kuzovlev97,Kogan01}  and when $\ell \ll \Lambda$, the characteristic length describing the spatial variation of the gauge-invariant phase across the junction is $\ell = \phi_0/4\pi \mu_0 \lambda^2 j_c,$
where $\phi_0 = h/2e$ is the superconducting flux quantum and $j_c$ (assumed to be independent of position) is the maximum Josephson current density that can flow as a supercurrent through the junction.  When $\ell \gg \Lambda$, the characteristic length scale is $\sqrt{\ell \Lambda}$.\cite{Kogan01} In this paper I assume that the junction length obeys both $W \ll \ell$ and $W \ll \sqrt{\ell \Lambda}$, such that the conditions are equivalent to the short-junction limit in sandwich-type Josephson junctions.\cite{Barone82,Orlando91}

The purpose of this paper is first to review how the screening current and the phase gradient induced in response to $\bm B$ affect $I_c(B)$ and then to calculate how  $I_c(B)$ is affected by the screening current and its phase gradient generated by a vortex or antivortex trapped near the junction.

\section{\label{phase}Gauge-invariant phase difference}

In the context of the Ginzburg-Landau (GL) theory,\cite{deGennes,StJames69} the superconducting order parameter can be expressed as $\psi = \psi_0 f e^{i\gamma}$, where $\psi_0$ is the magnitude of the order parameter in equilibrium, $f = |\psi|/\psi_0$ is the reduced order parameter, and $\gamma$ is the phase.  The second GL equation (in SI units) is 
\begin{equation}
\bm K = -\frac{2f^2}{\mu_0 \Lambda}(\bm A +\frac{\phi_0}{2\pi}\nabla \gamma),
\label{GL2}
\end{equation} 
where $\bm K = \bm j d$ is the sheet-current density, $\bm A$ is the vector potential, and  $\bm B = \nabla \times \bm A$ is the magnetic induction.  Since $\bm K$ is a gauge-invariant quantity, so is the quantity within the parentheses on the right-hand side. Different choices for the gauge of the vector potential $\bm A$ result in different expressions for $\gamma$.  

Consider the planar Josephson junction sketched in Fig.\ \ref{sample}.  With a sinusoidal current-phase relation, the Josephson current density in the $x$ direction across the junction of width $d_i$ at $x = 0$ is $K_x(y) = K_c \sin \Delta \gamma (y)$, where $K_c=j_c d$  is the maximum Josephson sheet-current density and $\Delta \gamma(y)$ is the gauge-invariant phase difference between the left (a) and right (b) superconductors,
\begin{equation}
\Delta \gamma(y) = \gamma_{\rm a}(-\frac{d_i}{2},y) - \gamma_{\rm b}(\frac{d_i}{2},y) -\frac{2\pi}{\phi_0}\int_{-d_i/2}^{d_i/2}
A_x(x,y) dx.  
\label{Deltagammadefinition}
\end{equation}

I assume here that the induced or applied sheet-current densities $\bm K_{\rm a}$ and $\bm K_{\rm b}$ on the left- and right-hand sides of the junction are so weak that the suppression of the magnitude of the superconducting order parameter  is negligible, such that $f = 1$.
A simple relation between these current densities and the gauge-invariant phase difference can be obtained by integrating the vector potential $\bm A$ around a very narrow rectangular loop of width $d_i$  in the $xy$ plane that just encloses the junction (with the bottom end at the origin and the top end at $y$), neglecting the magnetic flux up through the contour, and making use of Eq.\ (\ref{GL2}) with $f = 1$ for those portions of the integration along the sides of the junction:
\begin{equation}
\Delta \gamma(y) = \Delta \gamma_0+\frac{\pi \mu_0 \Lambda}{\phi_0}
\int_0^y [K_{{\rm b}y}(0,y')-K_{{\rm a}y}(0,y')]dy',
\label{Deltagammaintegral}
\end{equation} 
where $ \Delta \gamma_0 = \Delta \gamma(0)$.  In the presence of both an applied magnetic induction $B$ and trapped vortices, the sheet-current density in general is the vector sum of three contributions:\cite{Clem10,Clem10a} $\bm K = \bm K_J + \bm K_B + \bm K_v$, where $\bm K_J$ is generated by the injection of Josephson currents across the junction, $\bm K_B$ is induced by the applied magnetic induction $B$, and $\bm K_v$ is generated by the trapped vortices.  

The short-junction-limit assumption that  both $W \ll \ell$ and $W \ll \sqrt{\ell \Lambda}$ allows us to neglect the contributions from $\bm K_J$ on the right-hand side of Eq.\ (\ref{Deltagammaintegral}).\cite{Clem10,Clem10a}  Thus there are only two contributions to the sheet-current density and gauge-invariant phase difference we need to calculate: $\bm K_B$ and $\Delta \gamma_B$ induced by the applied magnetic induction and $\bm K_v$ and $\Delta \gamma_v$ generated by any nearby trapped vortices.  
 
\section{\label{DgB}$\Delta \gamma_B$ induced by an applied field WHEN NO VORTICES ARE TRAPPED NEARBY}

Let us first calculate the contributions to the sheet-current density $\bm K_B$ and the gauge-invariant phase difference $\Delta \gamma_B$ generated by a perpendicularly applied magnetic induction $\bm B = B \hat z$.  
Since  $\bm K_{B{\rm a}}$ and the corresponding phase field $\gamma_{B{\rm a}}$ easily can be obtained by symmetry from $\bm K_{B{\rm b}}$ and  $\gamma_{B{\rm b}}$, we can calculate only the latter in the region $x >0$ and suppress the subscript b.  

\begin{figure}
\includegraphics[width=8cm]{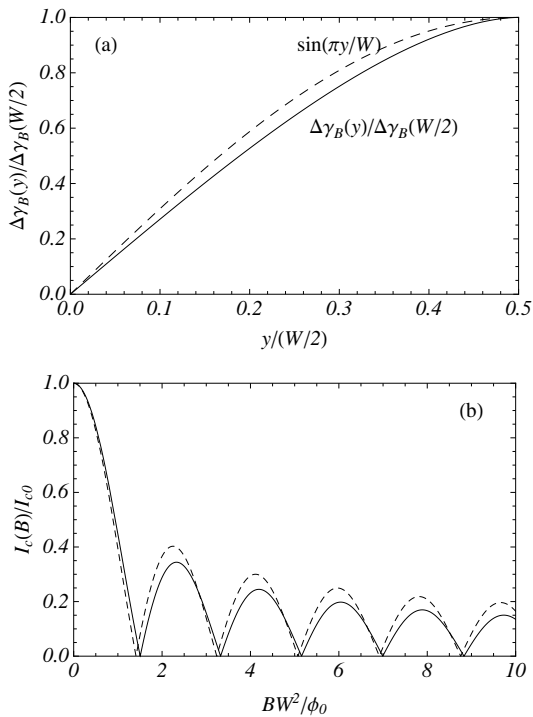}
\caption{%
(a) $\Delta \gamma_B (y)/\Delta \gamma_B (W/2)$ from Eq.\ (\ref{Deltagamma}) vs $y/(W/2)$ (solid curve).  For $y < 0$, note that $\Delta \gamma_B (-y) = -\Delta \gamma_B (y)$.  For comparison, the dashed curve shows $\sin (\pi y/W)$.
(b) $I_c(B)/I_{c0}$  from Eq.\ (\ref{Deltagamma})  vs $BW^2/\phi_0$ (solid curve). The dashed curve shows the Bessel-function approximation of Eq.\ (\ref{IcBBessel}).}
\label{DeltagammaIcplot}
\end{figure}

With the gauge choice $\bm A = -\hat x B y$, since $\nabla \cdot \bm K_B = 0$ [see Eq.\ (\ref{GL2})],  $\nabla^2 \gamma_B = 0$ must be solved subject to the boundary conditions following from  $K_{Bx}(0,y) =0$ and  $K_{By}(x,\pm W/2) = 0$, namely $\gamma_{Bx}(0,y)=2\pi B y/\phi_0$ and $\gamma_{By}(x, \pm W/2) = 0,$
 where $\gamma_{Bx} = \partial \gamma_B/\partial x$ and $\gamma_{By} = \partial \gamma_B/\partial y$.
The solution for $x > 0$, obtained by the method of separation of variables, is\cite{Clem10} (up to a constant)
\begin{eqnarray}
\gamma_B(x,y)\!\!\!& =& \!\!\!-\frac{8 B W^2}{\pi^2 \phi_0}\!\!\!\sum_{n=0}^\infty\!\!\frac{(-1)^n \!\exp[-(2n\!+\!1)X]\sin[(2n\!+\!1)Y] }{(2n+1)^3}\label{gamma1} \nonumber\\
&=&\frac{i B W^2}{2 \pi^2\phi_0}e^{-(X+iY)}[-\Phi(-e^{-2(X+iY)},3,1/2) \nonumber \\
&&+e^{2iY}\Phi(-e^{-2(X-iY)},3,1/2)],\label{gamma2}
\end{eqnarray}
where $X = \pi x/W$, $Y = \pi y/W$, and $\Phi(z,s,a)=\sum_{k=0}^\infty z^k/(k+a)^s$
is the Lerch transcendent.\cite{Mathematica} 
Since $\gamma_{\rm b}(d_i/2,y)$ in Eq.\ (\ref{Deltagammadefinition}) corresponds to $\gamma_B(0,y)$ and $\gamma_{\rm a}(-d_i/2,y) = -\gamma_B(0,y)$ by symmetry, 
the gauge-invariant phase difference given in Eq.\ (\ref{Deltagammadefinition}) can be obtained from Eq.\ (\ref{gamma2}) as
$\Delta \gamma_B (y)=-2\gamma_B(0,y)$:
\begin{eqnarray}
\Delta\gamma_B(y)\!\!\!& =& \!\!\!\frac{16 B W^2}{\pi^2 \phi_0}\!\sum_{n=0}^\infty\!\!\frac{(-1)^n \!\sin[(2n\!+\!1)Y] }{(2n+1)^3}\nonumber\\
&=&\frac{i B W^2}{\pi^2\phi_0}e^{-iY}[\Phi(-e^{-2iY)},3,1/2) \nonumber \\
&&-e^{2iY}\Phi(-e^{2iY)},3,1/2)].\label{Deltagamma}
\end{eqnarray}
The maximum value of $\Delta \gamma_B(y)$ occurs at $y = W/2$, where
\begin{equation}
\Delta \gamma_B(W/2)= \frac{14 \zeta(3) B W^2}{\pi^2 \phi_0}= 1.705 \frac{B W^2}{\phi_0}
\label{Deltagammamax}
\end{equation}
and $\zeta(3) = 1.20206$ is the Riemann zeta function.
Figure \ref{DeltagammaIcplot}(a) shows a plot of  $\Delta \gamma_B(y)/\Delta \gamma_B(W/2)$  vs $y/(W/2)$ and for comparison $\sin(\pi y/W)$ vs $y/(W/2)$. 
 If desired, the $x$ and $y$ components of the induced sheet-current density $\bm K_B(x,y)$ can  be obtained from Eqs.\ (\ref{GL2}) and (\ref{gamma2}).

The Josephson critical current $I_c(B)$, the maximum integral of $K_c \sin [\Delta \gamma_0+\Delta \gamma_B(y)]$ over $y$ from $-W/2$ to $W/2$, occurs when $\Delta \gamma_0 =\pm \pi/2$, such that 
\begin{equation}
\frac{I_c(B)}{I_{c0}} = \frac{1}{W}\Big|\int_{-W/2}^{W/2} 
\cos [\Delta \gamma_B(y)]dy \Big|,
\label{IcBnorm}
\end{equation}
where $\Delta \gamma_B(y)$ is given in Eq.\ (\ref{Deltagamma}) and $I_{c0} = K_cW$.  
The solid curve in Fig.\ \ref{DeltagammaIcplot}(b) shows a plot of $I_c(B)/I_{c0}$ vs $B W^2/\phi_0$. As noted in Ref.\ \onlinecite{Moshe08}, the maxima of $I_c(B)$ decrease as $1/\sqrt{B}$ instead of $1/B$ as in the familiar (Fraunhofer-like) bulk case.  Moreover, the spacings between the minima of $I_c(B)$ are not all the same, in contrast to the Fraunhofer pattern.

Let us define $\Delta B_1$ as the value of $B$ at which $I_c(B)$ has its first zero, $\Delta B_2$ as the difference of the values at which $I_c(B)$ has its second and first zeros, and $\Delta B_n$ as the difference of the values at which $I_c(B)$ has its $n$th and  $(n-1)$th zeros. 
For large $n$, the $\Delta B_n$ approach the limiting value
\begin{equation}
\Delta B = [\pi^3/14 \zeta (3)]\phi_0/W^2= 1.842\phi_0/W^2,
\label{DeltaB}
\end{equation}
as pointed out in Refs.\ \onlinecite{Rosenthal91} and \onlinecite{Humphreys93}.
The $\Delta B_n$ are smaller for small $n$ than for large $n$. Numerical evaluation of Eq.\ (\ref{IcBnorm}) yields the following values for $n$ = 1, 2, 3, 4, and 5:  $\Delta B_n/\Delta B$ = 0.8173, 0.9866,  0.9946, 0.9968, and 0.9979.  The first minimum of $I_c(B)$ occurs to the left or right of the origin $B=0$ at $\Delta B_1 = 1.505 \phi_0/W^2$, as can be seen in Fig.\  \ref{DeltagammaIcplot}(b).

If the $y$ dependence of $\Delta \gamma_B (y)$ is approximated by a sine function as in the dashed curve in Fig.\ \ref{DeltagammaIcplot}(a), then the integral in Eq.\ (\ref{IcBnorm}) can be evaluated in terms of the Bessel function $J_0$ with the result 
\begin{equation}
\frac{I_c(B)}{I_{c0}} = \Big|J_0\Big(\frac{14 \zeta(3) B W^2}{\pi^2 \phi_0}\Big)\Big|,
\label{IcBBessel}
\end{equation}
shown as the dashed curve in Fig.\ \ref{DeltagammaIcplot}(b).
For large $n$, the spacing between zeros for this approximation to $I_c(B)$ is exactly the same as in Eq.\ (\ref{DeltaB}), but from the well-known zeros of $J_0(x)$, we find the following values for $n$ = 1, 2, 3, 4, and 5:  $\Delta B_n/\Delta B$ = 0.7655, 0.9916,  0.9975, 0.9988, and 0.9993.

\section{\label{Dgv}$\Delta \gamma_v$ generated by a pinned vortex when no magnetic field is applied}

\begin{figure}
\includegraphics[width=7cm]{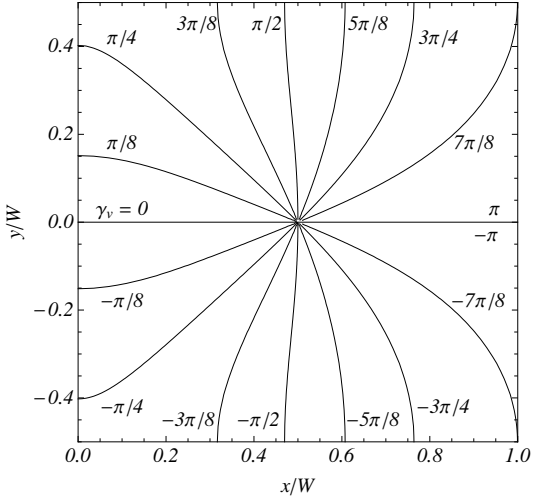}
\caption{%
Contour plot of the phase $\gamma_v(x_v,y_v;x,y)$ in the region $x >0$ around a Pearl vortex at $(x_v,y_v) = (0.5W,0)$.}
\label{gammav0p5plot}
\end{figure}

\begin{figure}
\includegraphics[width=7cm]{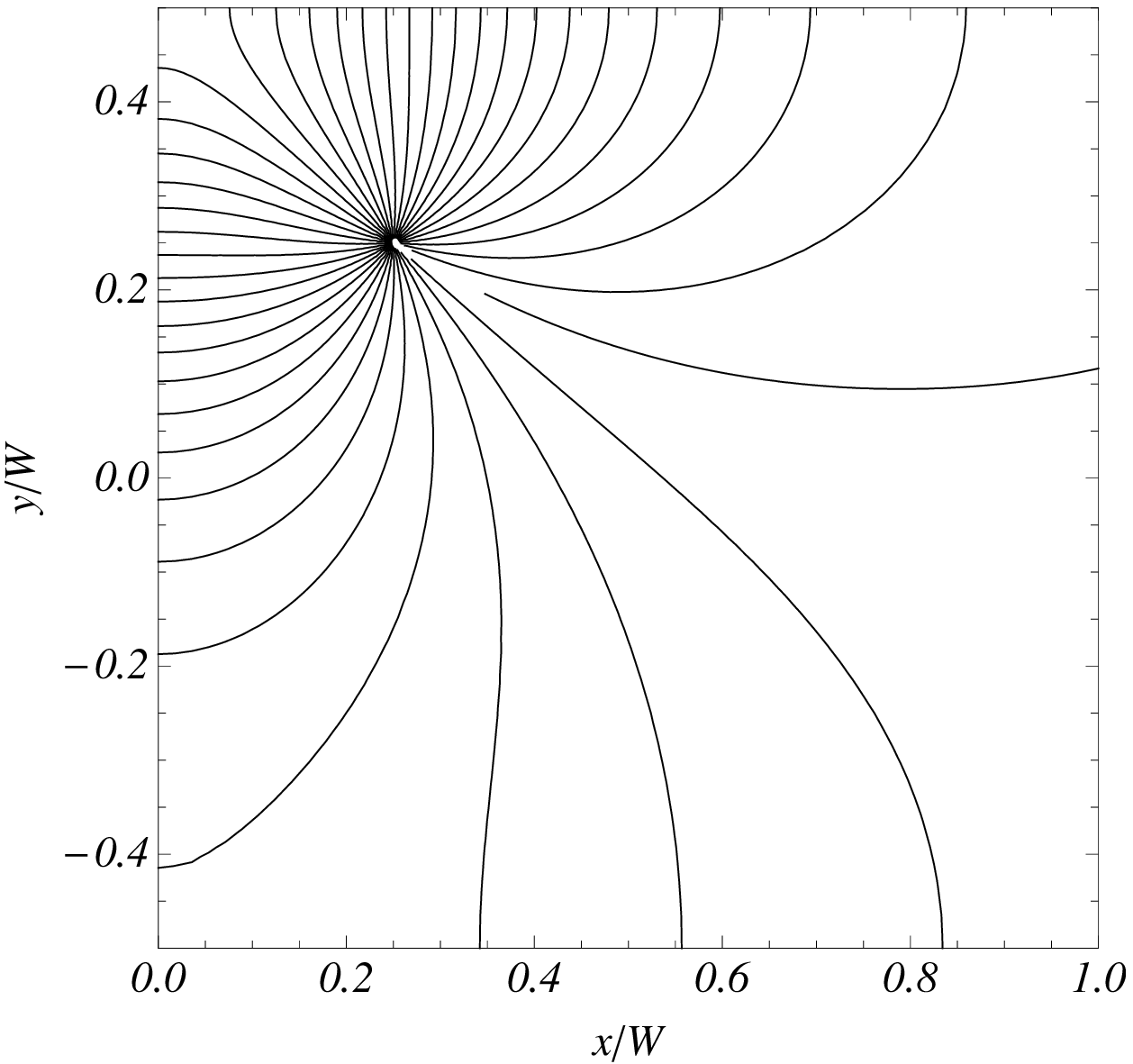}
\caption{%
Contour plot of the phase $\gamma_v(x_v,y_v;x,y)$ in the region $x >0$ around a Pearl vortex at $(x_v,y_v) = (0.25W,0.25W)$.}
\label{gammav0p250p25plot}
\end{figure}

Let us next calculate the contributions to the sheet-current density $\bm K_v$ and the gauge-invariant phase difference $\Delta \gamma_v$ generated by a $z$-oriented Pearl vortex centered at $(x_v,y_v)$ in side b, as shown in Fig.\ \ref{sample}.  
Since we are considering the limit $W \ll \Lambda$, we can ignore the magnetic field generated by the vortex but we must correctly account for the boundary conditions on the sheet-current density   $\bm K_{v{\rm b}}$ circulating around the vortex on side b.  Because $W \ll \Lambda$, the current density on side a is negligibly small ($\bm K_{v{\rm a}}= 0).$ Since we also may take the vector potential $\bm A$ to be negligibly small, Eq.\ (\ref{GL2}) and $\nabla \cdot \bm K_v = 0$ yield the equation $\nabla^2 \gamma_v = 0$, which must be solved subject to the boundary conditions following from  $K_{vx}(0,y) =0$ and  $K_{vy}(x,\pm W/2) = 0$: $\gamma_{vx}(0,y)=\gamma_{vy}(x, \pm W/2) = 0$,  where $\gamma_{vx} = \partial \gamma_v/\partial x$ and $\gamma_{vy} = \partial \gamma_v/\partial y$.  In addition, $\gamma_v$ must increase by $2\pi$ when traversing  a closed contour clockwise around the vortex axis: $\nabla \times \nabla \gamma_v = - \hat z 2\pi \delta(x-x_v)\delta(y-y_v)$.  The solution, obtained using conformal mapping,  is 
\begin{equation}
\gamma_v(x_v,y_v;x,y)= \Im \ln\Big(\frac{w(\zeta)-w^*(\zeta_v)}{w(\zeta_v)-w(\zeta)}\Big),
\label{gammav}
\end{equation}
 where $\Im$ denotes the imaginary part, $\zeta = x + i y$, $\zeta_v = x_v + i y_v$, and $w(\zeta) = i \sinh(\pi\zeta/W)$. 
Figure  \ref{gammav0p5plot} shows a plot of the vortex-generated phase when the vortex is at $(x_v,y_v) = (0.5W,0)$, and Fig.\ \ref{gammav0p250p25plot} shows a similar plot but with more contours for a vortex at $(x_v,y_v) = (0.25W,0.25W)$.
 If desired, the $x$ and $y$ components of the induced sheet-current density $\bm K_B(x,y)$ can  be obtained from Eqs.\ (\ref{GL2}) and (\ref{gammav}).

In the limit $W \ll \Lambda$ when there is no vortex on side a, $\gamma_{\rm a}(-d_i/2,y)=0$, while $\gamma_{\rm b}(d_i/2,y)=\gamma_v(x_v,y_v;0,y)$.  Since we can neglect the vector potential $\bm A$, Eq.\ (\ref{Deltagammadefinition}) thus yields the gauge-invariant phase difference  for a Pearl vortex (up to a constant) 
$\Delta \gamma_v(y) = -\gamma_v(x_v,y_v;0,y)$, i.e.,
\begin{equation}
\Delta \gamma_v(y) =-2\tan^{-1}\Big[\frac{\sin(\frac{\pi y}{W})-\cosh(\frac{\pi x_v}{W})\sin(\frac{\pi y_v}{W})}{\sinh(\frac{\pi x_v}{W})\cos(\frac{\pi y_v}{W})}\Big].
\label{Deltagammav}
\end{equation}
The sign of $\Delta \gamma_v(y)$ is reversed for a Pearl antivortex.

In zero applied field, the Josephson critical current $I_c$, the maximum integral of $K_c \sin [\Delta \gamma_0+\Delta \gamma_v(y)]$ over $y$ from $-W/2$ to $W/2$, occurs in general when $\tan \Delta \gamma_0 =\overline{\cos\Delta\gamma_v}/\overline{\sin\Delta\gamma_v}$, where
\begin{eqnarray}
\overline{\sin\Delta\gamma_v}&=&\frac{1}{W}\int_{-W/2}^{W/2} 
\sin [\Delta \gamma_v(y)]dy,\label{sinavg}\\
\overline{\cos\Delta\gamma_v}&=&\frac{1}{W}\int_{-W/2}^{W/2} 
\cos [\Delta \gamma_v(y)]dy,\label{cosavg}
\end{eqnarray}
such that, since $I_{c0} = K_c W$,
\begin{equation}
I_c= I_{c0} (\overline{\sin\Delta\gamma_v}^2+
\overline{\cos\Delta\gamma_v}^2)^{1/2}.
\label{Icv}
\end{equation}

\begin{figure}
\includegraphics[width=6cm]{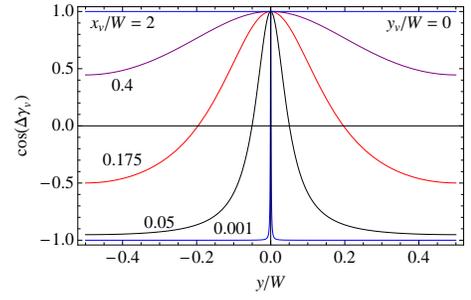}
\caption{%
(Color online)  Plot of $\cos[\Delta \gamma_v(y)]$ vs $y/W$ for a singly quantized Pearl vortex or antivortex on the $x$ axis ($y_v/W = 0$) for $x_v/W = 0.001$ (blue),  $0.05$ (black), $0.175$ (red), $0.4$ (purple), and $2$ (blue).}
\label{cosplot}
\end{figure}

\begin{figure}
\includegraphics[width=6cm]{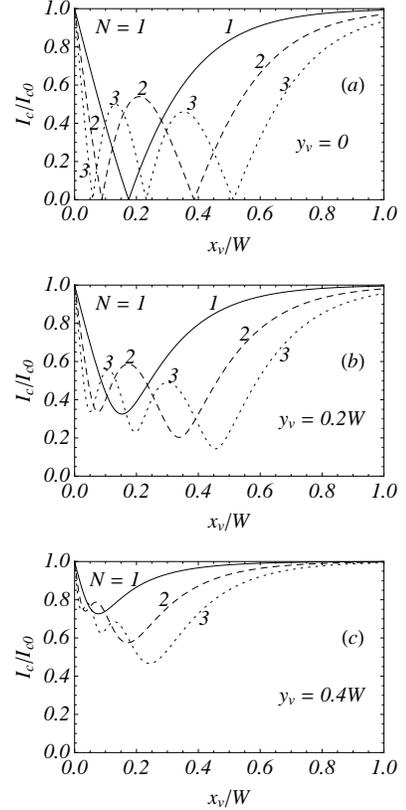}
\caption{%
Plot of the critical current for a singly ($N = 1$, solid), doubly ($N=2$, dashed), or triply ($N = 3$, dotted) quantized Pearl vortex or antivortex at $(x_v,y_v)$ as a function of its distance $x_v$ from the junction for (a) $y_v = 0$, (b) $0.2W$, and (c) $0.4W$.}
\label{Icvortexcolumnplot}
\end{figure}

When a singly quantized ($N=1$) vortex or antivortex is trapped on the $x$ axis, as in Fig.\ \ref{gammav0p5plot}, $\Delta\gamma_v(y)$ is an odd function of $y$, such that  $\overline{\sin\Delta\gamma_v}=0$, and 
\begin{equation}
\frac{I_c }{I_{c0}}=|\overline{\cos\Delta\gamma_v}| = \Big|1-2\tanh\Big(\frac{\pi x_v}{W}\Big)\Big|.
\end{equation}
Thus $I_c=0$ at one point along the $x_v$ axis, $x_v/W = \tanh^{-1}(1/2)/\pi = 0.175$.  The reason for this behavior is illustrated in Fig.\ \ref{cosplot}.  Note that $|\overline{\cos\Delta\gamma_v}| \approx 1$ for $x_v/W = 0.001$ and $2$, but that  $|\overline{\cos\Delta\gamma_v}| =0 $ for $x_v/W = 0.175.$   See also  Fig.\ \ref{Icvortexcolumnplot}(a) for $N=1$. 

Similarly, when a doubly quantized vortex (or antivortex) is trapped on the $x$ axis and the gauge-invariant phase $\Delta \gamma_v$ is doubled, 
\begin{equation}
\frac{I_c }{I_{c0}}= \Big|1-4\tanh\Big(\frac{\pi x_v}{W}\Big)\Big[1-\tanh^2\Big(\frac{\pi x_v}{W}\Big)\Big]\Big|,
\end{equation}
such that $I_c = 0$ at two points along the $x_v$ axis, $x_v/W = 0.088$ and 0.386 [see Fig.\ \ref{Icvortexcolumnplot}(a), $N=2$]. 
When a triply quantized vortex is trapped on the $x$ axis and the gauge-invariant phase $\Delta \gamma_v$ is tripled, 
\begin{eqnarray}
\frac{I_c }{I_{c0}}&=& \Big|1-2\tanh\Big(\frac{\pi x_v}{W}\Big) \nonumber \\
&\times&\!\!\!\!\Big[3-8\tanh^2\Big(\frac{\pi x_v}{W}\Big) +6\tanh^4\Big(\frac{\pi x_v}{W}\Big)\Big]\Big|,
\end{eqnarray}
such that $I_c = 0$ at three points along the $x_v$ axis,  $x_v/W = 0.059$, 0.232, and 0.513 [see Fig.\ \ref{Icvortexcolumnplot}(a), $N=3$]. 
However, when the vortex is trapped at a position off the $x$ axis, as in Fig.\ \ref{gammav0p250p25plot}, the zeros of $I_c$ are  replaced by minima.  As shown in  Fig.\ \ref{Icvortexcolumnplot}(b) and (c), $I_c$ vs $x_v$ exhibits one minimum for a singly quantized vortex ($N=1$), two for a doubly quantized vortex ($N=2$), and three for a triply quantized vortex ($N=3$).

\section{\label{Dgboth}$\Delta \gamma$ generated when a magnetic field is applied in the presence of a pinned vortex or antivortex}

We are now in a position to calculate how the $I_c(B)$ characteristics calculated in Sec.\ \ref{DgB} are affected by the presence of a vortex, described in Sec.\ \ref{Dgv}.  Since the resulting gauge-invariant phase difference $\Delta \gamma$ is (aside from a constant) simply the sum of the contributions $\Delta \gamma_B$ and $\Delta \gamma_v$ (and the resulting sheet-current density $\bm K$ is the sum of $\bm K_B$ and $\bm K_v$), the junction critical current in the presence of both an applied magnetic induction $\bm B = \hat z B$ and a $z$-directed Pearl vortex at $(x_v,y_v)$ is given by 
\begin{equation}
I_c(B)/I_{c0}= (\overline{\sin\Delta\gamma}^2+
\overline{\cos\Delta\gamma}^2)^{1/2},
\label{IcBboth}
\end{equation}
where $I_{c0} = K_cW$ and the averages are calculated as in Eqs.\ (\ref{sinavg}) and (\ref{cosavg}) but with $\Delta \gamma(y) = \Delta \gamma_B(y)+\Delta\gamma_v(y)$.  For the case of an applied magnetic induction and an antivortex, the sign of $\Delta \gamma_v(y)$ is reversed, and the averages are calculated with 
 $\Delta \gamma(y) = \Delta \gamma_B(y)-\Delta\gamma_v(y)$. For a vortex or an antivortex on the $x$ axis, since both $\Delta \gamma_B(y)$ and $\Delta\gamma_v(y)$ when $y_v = 0$ are odd functions of $y$, $\overline{\sin\Delta\gamma} = 0$, and $I_c(B)/I_{c0}= |\overline{\cos\Delta\gamma}|$.

\begin{figure}
\includegraphics[width=6.9cm]{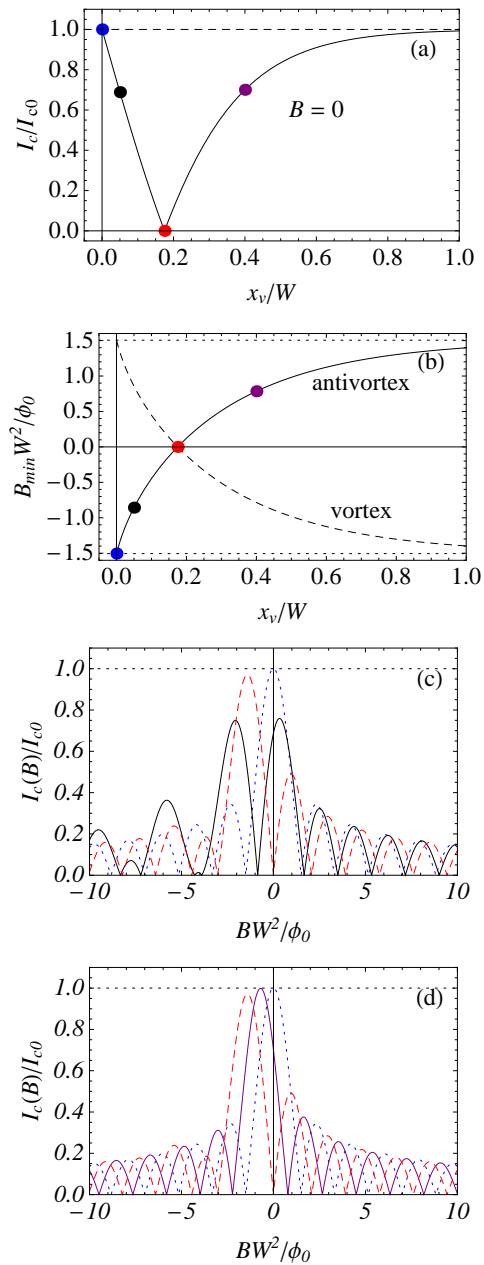}
\caption{%
(Color online) Behavior of the critical current for an antivortex at $x = x_v$ and $y = y_v = 0$:  (a) Normalized critical current $I_c/I_{c0}$ vs $x_v/W$ at $B=0$.  (b) Normalized position of the minimum of $I_c(B)$ nearest the origin, $B_{min}W^2/\phi_0$  vs $x_v/W$.  (c) Normalized critical current $I_c(B)/I_{c0}$ vs $BW^2/\phi_0$ for $x_v/W = 0$ (blue dotted curve), 0.05 (black solid curve), and 0.175 (red dashed curve.  (d) Normalized critical current $I_c(B)/I_{c0}$ vs $BW^2/\phi_0$ for $x_v/W = 0.175$ (red dashed curve), 0.4 (purple solid curve), and $\infty$ (blue dotted curve).}
\label{IcvortexvsBcolumnplot}
\end{figure}

Figure \ref{IcvortexvsBcolumnplot} exhibits the interesting behavior of how the pattern of $I_c(B)$ vs $B$ depends upon the position of a nearby Pearl antivortex at $x = x_v$ and $y = 0$.  First let us focus on the behavior of $I_c(0)$, which corresponds to the case discussed in Sec.\ \ref{Dgv}.  In the limit as $x_v/W \to 0$ [blue point in Fig.\ \ref{IcvortexvsBcolumnplot}(a)]  the critical current is simply $I_{c0} = K_c W$.  However, as the antivortex moves away from the junction, $I_c(0)$ drops to smaller values [black point at $x_v/W = 0.05$ in Fig.\ \ref{IcvortexvsBcolumnplot}(a)] and becomes zero  [red point  at $x_v/W = 0.175$ in Fig.\ \ref{IcvortexvsBcolumnplot}(a)].  As $x_v$ increases further, $I_c(0)$ rises [purple point at $x_v/W = 0.4$ in Fig.\ \ref{IcvortexvsBcolumnplot}(a)] and approaches $I_{c0} = K_cW$ as $x_v/W \to \infty$ as shown in Fig.\ \ref{IcvortexvsBcolumnplot}(a).

Next let us focus on the minimum at $B = B_{min}$, where initially in the limit  $x_v/W \to 0$ [blue point  at $x_v/W = 0$ in Fig.\ \ref{IcvortexvsBcolumnplot}(b) and blue dotted curve in Fig.\ \ref{IcvortexvsBcolumnplot}(c)]  $B_{min}W^2/\phi_0 = -1.505$,  to the left of the origin of Fig.\   \ref{IcvortexvsBcolumnplot}(c).  As the antivortex moves away from the junction, this minimum moves to the right, as shown by the black point  at $x_v/W = 0.05$ in Fig.\ \ref{IcvortexvsBcolumnplot}(b) and the black solid curve in Fig.\ \ref{IcvortexvsBcolumnplot}(c).  When the antivortex reaches the point $x_v/W = 0.175$ [red point in Fig.\ \ref{IcvortexvsBcolumnplot}(b) and red dashed curve in Fig.\ \ref{IcvortexvsBcolumnplot}(c) and (d)], the minimum occurs at $B = 0$. As $x_v$ increases further, this minimum continues to move to the right [purple point at $x_v/W = 0.4$ in Fig.\ \ref{IcvortexvsBcolumnplot}(b) and purple solid curve in Fig.\ \ref{IcvortexvsBcolumnplot}(d)].  As  $x_v/W \to \infty$, the minimum occurs at 
$B_{min}W^2/\phi_0 = +1.505$, as shown in Fig.\ \ref{IcvortexvsBcolumnplot}(b), and the pattern of $I_c(B)/I_{c0}$ becomes the same as in the absence of a vortex or antivortex [blue dotted curve in Fig.\ \ref{IcvortexvsBcolumnplot}(d) or black solid curve in Fig.\ \ref{DeltagammaIcplot}(b)].  

Now let us examine the behavior of the pattern of primary and secondary maxima in $I_c(B)/I_{c0}$ as $x_v$ increases.  When $x_v/W = 0$, there is a secondary maximum of height 0.34 at $BW^2/\phi_0 = -2.33$ and a primary maximum of height 1.00 at $BW^2/\phi_0 = 0$ [blue dotted curve in Fig.\ \ref{IcvortexvsBcolumnplot}(c)].  As $x_v/W$ increases, the secondary maximum moves to the right and grows in height until it becomes the primary maximum, while the primary maximum also moves to the right but decreases in height until it becomes a secondary maximum.  These changes can be seen in Fig.\ \ref{IcvortexvsBcolumnplot} in the progression of the black solid curve in (c), the red dashed curve in (c) and (d), the purple solid curve in (d), and the blue dotted curve in (d).  

Finally we note the asymmetry of the pattern of the primary and secondary maxima as $x_v $ increases.  Although $I_c(B)/I_{c0}$ has mirror symmetry in the limits $x_v/W \to 0$ and $x_v/W \to \infty$, as shown by the blue dotted curves in Fig.\ \ref{IcvortexvsBcolumnplot}(c) and (d), this symmetry is broken for intermediate values of $x_v/W$.  The asymmetry is most pronounced for $0 < x_v/W \le 0.175$.  For example, for $x_v/W = 0.05$ [black solid curve in   Fig.\ \ref{IcvortexvsBcolumnplot}(c)], the secondary maxima  decrease monotonically for increasing $B > 0$ and have nearly the same period.  However, for negative $B$ the heights of the secondary maxima are irregular, and the larger secondary maxima have approximately double the period of those for $B > 0$.  These effects also occur for $x_v/W = 0.175$ [red dashed curve in   Fig.\ \ref{IcvortexvsBcolumnplot}(c) and (d)] but are less pronounced.  For $x_v/W = 0.4$ [purple curve in   Fig.\ \ref{IcvortexvsBcolumnplot}(d)], asymmetry is present, with the secondary peaks for $B < 0$ lower than those for $B > 0$, but the period doubling is no longer present.

\begin{figure}
\includegraphics[width=7cm]{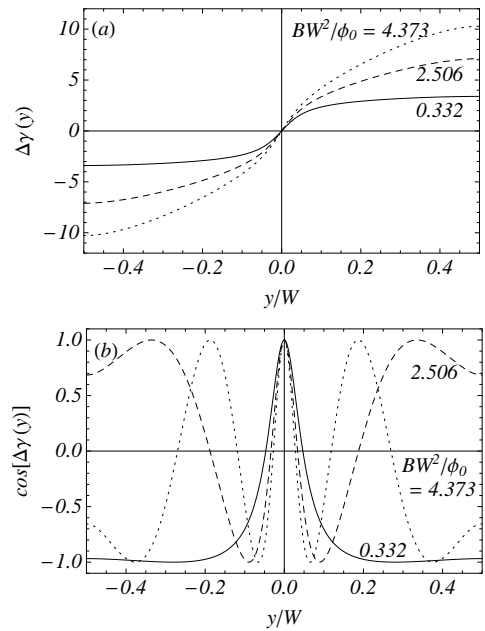}
\caption{%
(a) $\Delta \gamma(y)$ and (b) $\cos[\Delta \gamma(y)]$ when $x_v/W = 0.05$ and $y_v = 0$ [black solid curve in Fig.\ \ref{IcvortexvsBcolumnplot}(c)] for the first three maxima  for positive $B$, for which $BW^2/\phi_0$ and $I_c(B)/I_{c0}$ are 0.332 and 0.759; 2.506 and 0.323; and 4.373 and 0.236, respectively.}
\label{PositiveBplot}
\end{figure}

\begin{figure}
\includegraphics[width=7cm]{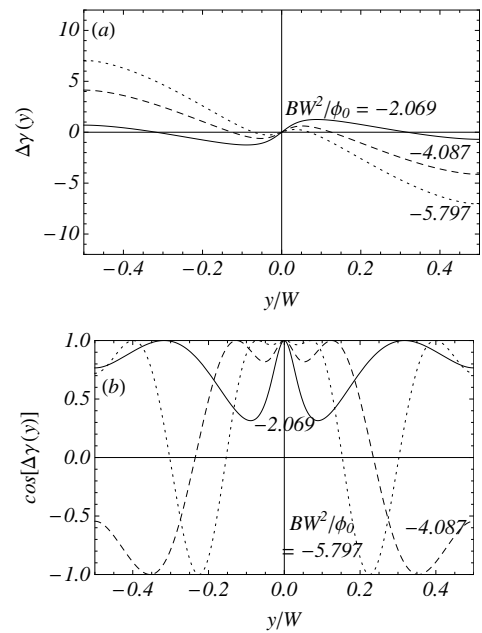}
\caption{%
(a) $\Delta \gamma(y)$ and (b) $\cos[\Delta \gamma(y)]$ when $x_v/W = 0.05$ and $y_v = 0$ [black solid curve in Fig.\ \ref{IcvortexvsBcolumnplot}(c)] for the first three maxima  for negative $B$, for which $BW^2/\phi_0$ and $I_c(B)/I_{c0}$ are -2.069 and 0.750; -4.087 and 0.013; and -5.797 and 0.363, respectively.}
\label{NegativeBplot}
\end{figure}The basic reason for the asymmetry of $I_c(B)$ about $B = 0$ when $y_v = 0$ is that $I_c(B)/I_{c0}= |\overline{\cos\Delta\gamma}|$, where 
 $\Delta \gamma(y) = \Delta \gamma_B(y)-\Delta\gamma_v(y)$.
Although $\Delta \gamma_B(y)$ is an antisymmetric function of $B$, $-\Delta\gamma_v(y)$ is independent of $B$, so that when $B \ne 0$, $\Delta \gamma(y)$ is neither symmetric nor antisymmetric about $B=0$. The effects of this asymmetry can be very pronounced, as seen in the example of the black solid curve in   Fig.\ \ref{IcvortexvsBcolumnplot}(c). Plots of $\Delta \gamma(y)$ and $\cos[\Delta \gamma(y)]$ calculated for $x_v/W = 0.05$, shown in Fig.\ \ref{PositiveBplot} at the first three maxima for $B > 0$ and Fig.\ \ref{NegativeBplot} at the first three maxima for $B < 0$, show the dramatic differences responsible for the asymmetry of $I_c(B)/I_{c0}$ and the approximate period doubling for $B < 0$.

Numerical calculations of how the $I_c(B)$ vs $B$ patterns for a doubly ($N=2$) or triply ($N=2$) quantized antivortex on the $x$ axis evolve as $x_v$ increases from zero to values of order $W$ or larger reveal behavior similar to those for a singly ($N=1$) quantized antivortex shown in Fig.\  \ref{IcvortexvsBcolumnplot}.  As $x_v$ increases, the patterns shift to the right, and for intermediate values of $x_v$ the maxima decrease monotonically for $B >0$ but have irregular heights for $B < 0$.  The chief difference from the behavior for $N=1$ is that  $I_c(0)$ passes through zero twice for $N = 2$ and three times for $N = 3$ in accordance with Fig.\ \ref{Icvortexcolumnplot}(a).

\section{\label{summary}Summary}

In this paper I first reviewed how the gauge-invariant phase difference $\Delta \gamma (y)$ across a planar Josephson junction in a long, thin superconducting film is affected by the sheet-current distributions on opposite sides of the junction.  The assumptions that  $W \ll \Lambda$ and $W \ll \ell$ made it possible to calculate the two relevant independent contributions to the gauge-invariant phase difference: $\Delta \gamma_B(y)$ due to the perpendicular applied magnetic induction $B$ and $\Delta \gamma_v(y)$ due to a nearby trapped Pearl vortex or antivortex.  After calculating the critical current $I_c$ of the junction for these two contributions separately, I calculated $I_c(B)$ when both $B$ is applied and a vortex or antivortex is near the junction.  

\begin{figure}
\includegraphics[width=6.5cm]{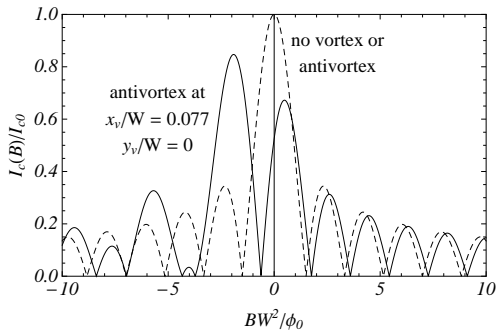}
\caption{%
Theoretically calculated normalized critical current $I_c(B)/I_{c0}$ vs $BW^2/\phi_0$ in the absence of a vortex or antivortex (dashed curve) and in the presence of an antivortex at $x_v/W = 0.077$ and $y = y_v  =0$ (solid curve).}
\label{Icsimplot}
\end{figure}

\begin{figure}
\includegraphics[width=6.5cm]{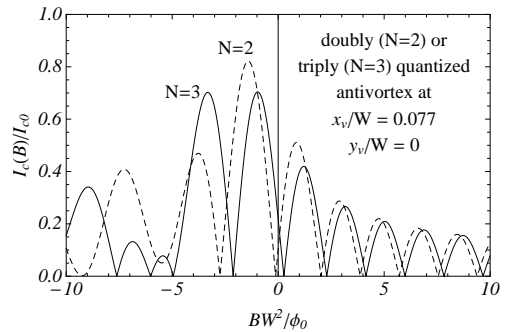}
\caption{%
Theoretically calculated normalized critical current $I_c(B)/I_{c0}$ vs $BW^2/\phi_0$ in the presence of a doubly quantized antivortex ($N=2$, dashed) and a triply quantized antivortex ($N=3$, solid) at $x_v/W = 0.077$ and $y = y_v  =0$.}
\label{IcN=2&3}
\end{figure}

The features observed in the calculated $I_c(B)$ vs $B$ characteristics show many features in common with the experimental $I_c(B)$ vs $B$ characteristics observed recently by Golod et al.\cite{Golod10}  The dashed curve in Fig.\ \ref{Icsimplot} shows the calculated curve of $I_c(B)$ vs $B$ in the absence of a vortex or antivortex, which shows perfect mirror symmetry about $B = 0$, a primary maximum at $B = 0$ and secondary maxima of monotonically decreasing heights for increasing $|B|$.  The corresponding experimental curve [Fig.\ 3(a) in Ref.\ \onlinecite{Golod10}] shows approximate mirror symmetry about $B=0$, a primary  maximum at $B =0$ and secondary maxima, which generally decrease in height for increasing $|B|$ but not monotonically.  The spacings of the minima and maxima along the $B$ axis increase for increasing $|B|$, as expected from the discussion in the paragraph containing Eq.\ (\ref{DeltaB}).  

Figure 3(b) in Ref.\ \onlinecite{Golod10} showed $I_c(B)$ vs $B$ for one antivortex trapped in a hole at a distance 0.29 $\mu$m  from the center of the junction, whose length was  3.8 $\mu$m.  In our model of the experiment, this corresponds to having an antivortex at $x_v/W = 0.29/3.8 = 0.077$ and $y_v = 0$, for which the calculated $I_c(B)$ vs $B$ is shown by the solid curve in Fig.\ \ref{Icsimplot}. This curve shows a primary maximum shifted to the left of $B = 0$ with secondary maxima to the right of the primary maximum monotonically decreasing for increasing $B$, and secondary maxima of irregular heights to the left of the primary maximum showing an approximate period doubling.  The experimental plot of $I_c(B)$ vs $B$, shown in Fig.\ 3(b) in Ref.\ \onlinecite{Golod10}, exhibits similar features:  a primary maximum shifted to the left of $B = 0$, secondary maxima to the right of the primary maximum monotonically decreasing for increasing $B$, and secondary maxima of irregular heights to the left of the primary maximum showing an approximate period doubling. However, the experimental  $I_c(B)$ showed a minimum at $B = 0$, while the theoretical curve has this minimum shifted to the left of the origin.  

Figure \ref{IcN=2&3} shows theoretical predictions of $I_c(B)$ vs $B$ patterns for a doubly quantized antivortex ($N=2$, dashed) or a triply quantized antivortex ($N = 3$, solid) trapped in the hole at $x_v/W = 0.077$ and $y_v = 0$.  Note the monotonic decrease of the maxima for $B > 0$ and irregular heights for $B < 0$.

Although the theoretical model does not assume the exact geometry and material properties of the sample used in Ref.\ \onlinecite{Golod10}, the theoretical results presented here provide a good qualitative and semi-quantitative explanation of the experimental results.

\begin{acknowledgments}
I thank V. M. Krasnov, J. E. Sadleir, and V. G. Kogan for stimulating discussions and T. Golod for helpful correspondence.
This research, supported by the U.S. Department of
Energy, Office of Basic Energy Science, Division of Materials
Sciences and Engineering, was performed at
the Ames Laboratory, which is operated for the U.S. Department
of Energy by Iowa State University under Contract No.
DE-AC02-07CH11358.  

\end{acknowledgments}


\begin{thebibliography}{99}
\bibitem{Golod10} T. Golod, A. Rydh, and V. M. Krasnov, \prl {\bf 104}, 227003 (2010).
\bibitem{Pearl64} J. Pearl, \apl {\bf 5}, 65 (1964). 
\bibitem{Moshe08} M. Moshe, V. G. Kogan, and R. G. Mints, \prb {\bf 78}, 020510(R) (2008).
\bibitem{Clem10} J. R. Clem, \prb {\bf 81}, 144515 (2010).
\bibitem{Ivanchenko90} Yu. M. Ivanchenko and T. K. Soboleva, Phys. Lett. A {\bf 147}, 65 (1990).
\bibitem{Gurevich92} A. Gurevich, \prb {\bf 46}, 3187 (1992).
\bibitem{Mints94} R. G. Mints and I. B. Snapiro, \prb {\bf 49}, 6188 (1994).
\bibitem{Ivanchenko95} Yu. M. Ivanchenko, \prb {\bf 52}, 79 (1995).
\bibitem{Kuzovlev97} Yu. E. Kuzovlev and A. I. Lomtev, Zh. Eksp. Teor. Fiz {\bf 111}, 1803 (1997) [JETP {\bf 84}, 986 (1997)].
\bibitem{Kogan01} 	V. G. Kogan, V. V. Dobrovitski, J. R. Clem, Y. Mawatari, and R. G. Mints, \prb {\bf 63}, 144501 (2001).
\bibitem{Barone82} A. Barone and G. Paterno, {\it Physics and Applications of the Josephson Effect}, (Wiley, New York, 1982).
\bibitem{Orlando91} T. P. Orlando and K. A. Delin, {\it Foundations of Applied Superconductivity}, (Addison-Wesley, Reading, 1991).
\bibitem{deGennes} P. G. de Gennes, {\it Superconductivity of Metals and Alloys} (Benjamin, New York, 1966), p. 177.
\bibitem{StJames69} D. Saint-James, E. J. Thomas, and G. Sarma, {\it Type II Superconductivity} (Pergamon, Oxford, 1969).
\bibitem{Clem10a} J. R. Clem, \prb {\bf 82}, 174515 (2010).
\bibitem{Mathematica} Wolfram Research, Inc., Mathematica, Version 7.0, Champaign, IL (2008).
\bibitem{Rosenthal91} P. A. Rosenthal, M. R. Beasley, K. Char, M. S. Colclough, and G. Zaharchuk, \apl {\bf 59}, 3482 (1991).
\bibitem{Humphreys93}  R. G. Humphreys and J. A. Edwards, Physica C {\bf 210}, 42 (1993).
\end{thebibliography}
\end{document}